\documentclass[aps,prl,reprint,superscriptaddress,amsmath,amssymb,showpacs]{revtex4-1}

\usepackage{graphicx}

\begin{document}

\title{Origins of shortcomings in recent realistic multiband Monte-Carlo studies for Ga$_{1-x}$Mn$_{x}$As}

\author{Stefan Barthel}
\email[]{sbarthel@itp.uni-bremen.de}
\affiliation{Institute for Theoretical Physics, University of Bremen, Otto-Hahn-Allee 1, D-28359 Bremen, Germany}

\author{Gerd Czycholl}
\affiliation{Institute for Theoretical Physics, University of Bremen, Otto-Hahn-Allee 1, D-28359 Bremen, Germany}

\author{Georges Bouzerar}

\affiliation{Institut N\'eel, 25 avenue des Martyrs, B.P. 166, 38042 Grenoble Cedex 09, France}
\affiliation{School of Engineering and Science, Jacobs University Bremen, Campus Ring 1, D-28759 Bremen, Germany}

\date{May 5, 2011}

\begin{abstract}
Magnetic properties of Mn-doped GaAs are re-investigated within a realistic multiband description of the host valence bands.
We explicitely demonstrate that the recent Monte Carlo (MC) simulations performed on a large scale supercomputer suffer from severe shortcomings.
Indeed, it is shown, using identical parameters that (i) the calculated Zeeman-splitting largely underestimates that obtained from first principle studies, (ii) the couplings exhibit strong RKKY oscillations, (iii) the stability region for ferromagnetism is much narrower than obtained previously and (iv) the calculated Curie temperatures appear to be at least one order of magnitudes smaller. We show that the proposed choice of physical parameters cannot describe the physics in (Ga,Mn)As.
\end{abstract}

%\pacs{71.10.Fd, 71,70.Gm, 75.30.Hx, 75.40.Mg, 75.50.Pp}

\maketitle

\section{Introduction}
In the recent years, dilute magnetic semiconductors (DMS) have risen a lot of interest because of their possible use for spintronic devices \cite{RevModPhys.76.323} (e.g. a spin-polarized light-emitting diode, etc.). A key goal is the achievement of a Curie temperature T$_C$ well above room temperature and a proper theoretical description of the involved physics was mainly controversial. From the theoretical point of view there are mainly two different kinds of approaches, (i) realistic bandstructure model studies ($\textbf{k}\!\!\cdot\!\!\textbf{p}$, Kohn-Luttinger, empirical tight-binding) \cite{RevModPhys.78.809} and (ii) sophisticated material specific ab-initio based (LSDA, SIC-LDA, ...) calculations \cite{RevModPhys.82.1633}. Note that the prediction of a large T$_C\approx700$ K in GaMnN \cite{Dietl11022000},  has triggered a considerable amount of both experimental and theoretical work. Later it has been demonstrated that this huge T$_C$ results from several drastic approximations namely (1) perturbative treatment, (2) mean-field (MF) and (3) virtual crystal approximation for disorder \cite{PhysRevLett.93.137202,0295-5075-69-5-812}. 
Recently it was shown that the essential properties of DMS can already be described within a single-band model \cite{0295-5075-92-4-47006,1367-2630-13-2-023002}. Therefore a model with a realistic bandstructure and a non-perturbative coupling between carriers and magnetic impurities should provide a powerful tool to study both transport and magnetic properties in the wide family of DMS.

Recently such a study was performed for the case of GaMnAs using large scale Monte-Carlo (MC) simulations \cite{popescu_crossover_2007,yildirim_large-scale_2007,moreo_multi-orbital_2007}. Although a MC approach can be considered as essentially exact, the study of dilute magnetic systems of classical spins coupled to quantum carriers should be done in a very careful way. Our intent is to discuss and compare our calculations based on a two step-approach (TSA) to the MC simulations of Ref.\cite{yildirim_large-scale_2007}. As it will be shown several crucial features are incorrect in Ref.\cite{yildirim_large-scale_2007} and explanations will be provided.
\section{Model}
We start from the following V-J model Hamiltonian
\begin{equation}
\hat{H} = \sum_{ij,\alpha\beta,\sigma}t_{ij}^{\alpha\beta}
\hat{c}_{i\alpha\sigma}^{\dagger}\hat{c}_{j\beta\sigma}
 + J_{pd}\sum_{i}p_i\hat{\bf{S}}_i\hat{\bf{s}}_i
 + V\sum_{i,\alpha,\sigma}p_i\hat{n}_{i\alpha\sigma}.
\label{eq:hamiltonian}
\end{equation}
The first term provides a realistic multiband description of the host valence bands.
Here $\hat{c}_{i\alpha\sigma}^{\dagger}$ $(\hat{c}_{i\alpha\sigma})$ is the creation (annihilation) operator of a carrier
with spin $\sigma$ at lattice site i. $\alpha, \beta$ denote the three different p-orbitals. The hopping matrix elements $t_{ij}^{\alpha\beta}$ are restricted to nearest neighbor
sites and are determined from known Luttinger parameters $\gamma_i$ \cite{PhysRev.102.1030}. We used $\gamma_{1,2,3}=\{6.85,2.1,2.9\}$ as used in Ref.\cite{yildirim_large-scale_2007} for GaAs.  The second term corresponds to the interaction between itinerant carrier spins $\hat{\bf{s}}_i$ and localized impurity spins $\hat{\bf{S}}_i$ (S = 5/2 for Mn$^{2+}$).
J$_{pd}$ is the local p-d coupling. The last term (on-site impurity potential scattering term) will be neglected (V = 0) as done in Ref.\cite{yildirim_large-scale_2007}. The relevance of the on-site term will be discussed in what follows. The parameter $p_i$ takes the values
$p_i=1$ if the site i is an impurity site otherwise $p_i=0$.
Now the general procedure is as follows:
First we calculate the total spectrum of the Hamiltonian defined in
Eq.($\ref{eq:hamiltonian}$)
for a given set of disorder configurations assuming the Mn$^{2+}$
impurity spins to be fully polarized. Then the Mn-Mn effective exchange
couplings $J_{ij}$
(see Ref.\cite{Lichtenstein198765,katsnelson_first-principles_2000}
for details) between two impurities located at
$\textbf{r}_i$ and $\textbf{r}_j$ are obtained by the relation:
\begin{equation}
J_{ij}=\frac{1}{4\pi S^2}\mathfrak{Im}\int\limits_{-\infty}^{\infty}f(\omega)
\mathrm{Tr}_{\alpha}\{\hat{\Sigma}_{i}\hat{G}_{ij}^{\uparrow}(\omega)
\hat{\Sigma}_{j}\hat{G}_{ji}^{\downarrow}(\omega)\}\mathrm{d}\omega.
\label{eq:couplings}
\end{equation}
Note that Eq.(\ref{eq:couplings}) depends on the
impurity concentration $x$ and on the hole density p.
In our model the self-energy reduces to
$\hat{\Sigma}_{i}$ = J$_{pd}$S$\cdot\mathrm{\hat{1}}$ and $\hat{G}^{\sigma}_{ij}(\omega)$ is the Green's function
of the system while the trace is taken with
respect to the orbitals.
$f(\omega)={\left( e^{\beta(\omega-E_{F})}+1\right)}^{-1}$ is the Fermi function and E$_{F}$ is the Fermi energy.
Finally the disordered Heisenberg model,
\begin{equation}
\hat{H} = -\sum_{i\neq j}J_{ij}p_ip_j\hat{\bf{S}}_i\cdot\hat{\bf{S}}_j,
\label{eq:heisenberg}
\end{equation}
defined
with the effective exchange couplings of
Eq.(\ref{eq:couplings}) is solved within
the self-consistent local random phase approximation (SC-LRPA).
This method \cite{0295-5075-69-5-812,PhysRevB.81.172406,0295-5075-92-4-47006}
has been proven to be a reliable tool to calculate e.g. Curie temperatures,
the magnon excitation spectrum and optical conductivity. The TSA was able to reproduce both ab-initio results and experimental data.
\section{Results}
\begin{figure}
 \includegraphics[scale=0.19]{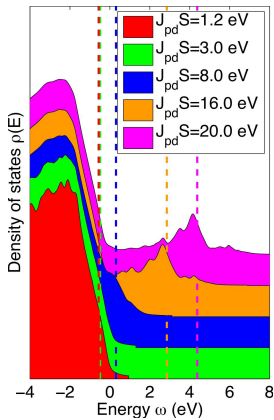}
 \caption{\label{fig:1}Empirical tight-binding (ETBM) density of states for
Ga$_{1-x}$Mn$_{x}$As for $x=0.085$ and different values
of J$_{pd}$S.
The dashed lines indicate the chemical potentials for $T = 0$ K corresponding to the hole density p = 0.75$x$.}
\end{figure}

\begin{figure}
\includegraphics[scale=0.35]{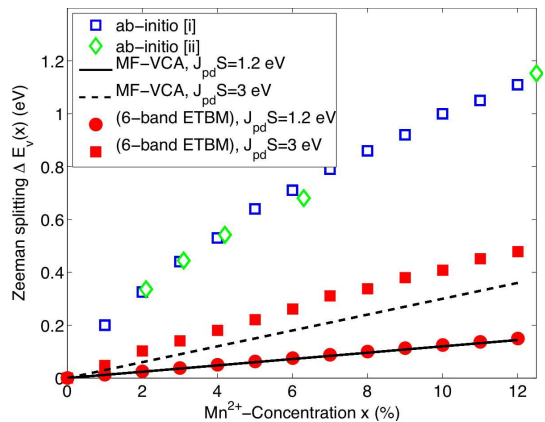}
\caption{\label{fig:2}Zeeman splitting $\Delta E_v(x)$ as a function of the Mn
concentration $x$ for Ga$_{1-x}$Mn$_{x}$As calculated within LSDA [i]\cite{kudrnovsky_private} [ii]\cite{wierzbowska_different_2004} and within the V-J model.}
\end{figure}

\begin{figure}
 \includegraphics[scale=0.35]{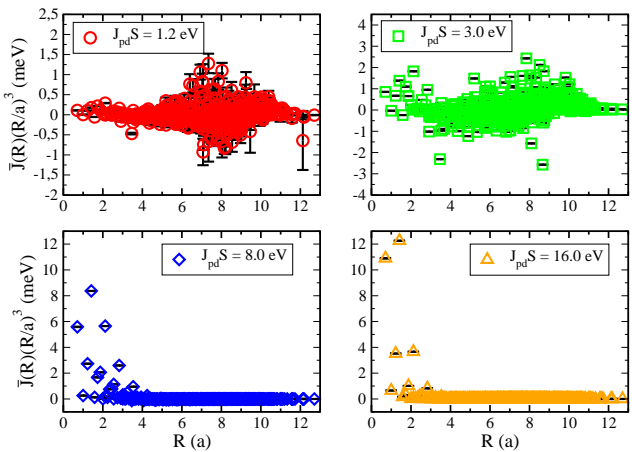}
 \caption{\label{fig:3}Mean effective exchange couplings $\bar{J}(R)\cdot R^3$ as a function of distance $R$ ($a$ is the lattice constant)
for several values of J$_{pd}$S and p = 0.75$x$ using $x = 0.085$. Please note the oscillating RKKY nature of the couplings for
J$_{pd}$S = $\{1.2, 3\}$ eV in contrast to the short-range ferromagnetic couplings
for J$_{pd}$S = $\{8,16\}$ eV.}
\end{figure}

\begin{figure}
\includegraphics[scale=0.35]{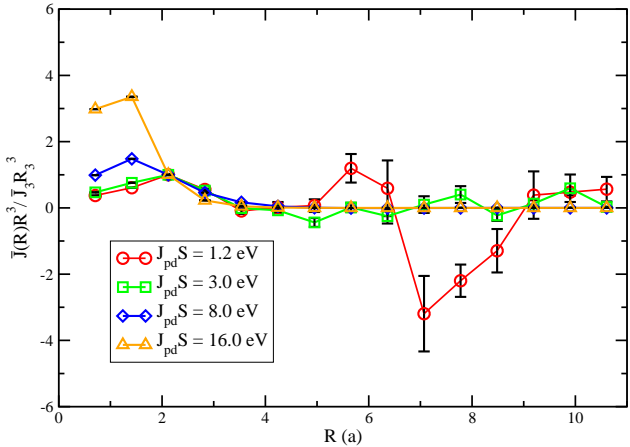}
\caption{\label{fig:4}Mean effective exchange couplings $\bar{J}(R)\cdot R^3$/($\bar{J}_3\cdot R_3^3)$ as a function of distance $R$
for several values of J$_{pd}$S and p=0.75$x$ using $x = 0.085$ in the [110] direction.}
\end{figure}

\begin{figure}
\includegraphics[scale=0.35]{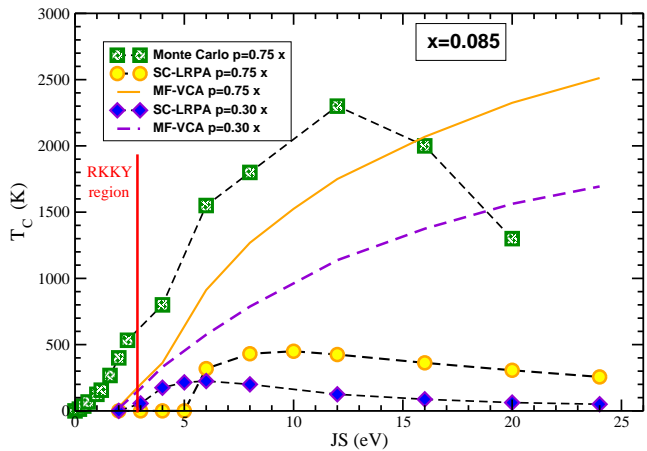}
\caption{\label{fig:5}Calculated Curie temperature T$_C$ for Ga$_{1-x}$Mn$_{x}$As with $x = 0.085$ using the V-J model in connection with V = 0. The SC-LRPA calculations were done with a system of 4$\cdot$(22)$^3$ sites (3600 impurities) and the average was performed over 100 disorder configurations.}
\end{figure}

For a fixed Mn impurity concentration of $x=0.085$ in Ga$_{1-x}$Mn$_{x}$As
the density of states (DOS) is shown in FIG. \ref{fig:1}. for
different values of the effective p-d coupling strength J$_{pd}$S.
These results are in perfect agreement with Ref.\cite{yildirim_large-scale_2007}. The DOS is almost unchanged compared to that of the pure system for J$_{pd}$S $\le3$ eV. This already gives a first indication/hint that this range of parameters corresponds to the perturbative RKKY regime, this will be confirmed in what follows. Let us proceed further by calculating the Zeeman splitting
$\Delta E_v(x)=E^{\uparrow}_{max}-E^{\downarrow}_{max}$ ($E^{\sigma}_{max}$
is the largest eigenvalue in the corresponding $\sigma$
sector) as a function of the Mn concentration. It is depicted in FIG. \ref{fig:2}. together with available ab-initio results.
For J$_{pd}$S = 1.2 eV $\Delta E_v$ is found to be in perfect agreement with the mean-field expression $\Delta E_v^{MF}(x)=x$J$_{pd}$S, the accordance is still reasonable
for J$_{pd}$S = 3 eV. Remark that, it is now widely accepted that J$_{pd}$ $\approx$ 1.2 eV in Mn-doped GaAs \cite{PhysRevB.58.R4211,Bhattacharjee199917}. Thus the realistic value of J$_{pd}$S should be 3 eV since S = 5/2 for Mn$^{2+}$. However in Ref.\cite{yildirim_large-scale_2007} J$_{pd}$S = 1.2 eV was used, which is almost three times smaller than the realistic value \cite{PhysRevLett.100.229701,PhysRevLett.100.229702}. From FIG. \ref{fig:2}. one can clearly see that for both J$_{pd}$S = $\{1.2,3\}$ eV the calculated Zeeman-splitting largely underestimates that obtained from first-principle studies \cite{wierzbowska_different_2004,kudrnovsky_private}. For example, for $x=0.05$, one finds
$\Delta E_v=0.06$ and 0.2 eV for J$_{pd}$ = 1.2 eV and 3 eV respectively, in contrast to 0.65 eV obtained from LSDA calculations. Thus, even the correct value of J$_{pd}$S does not lead to an agreement with first-principle results. In addition, experimental studies \cite{PhysRevLett.18.443,PhysRevB.55.6938,Yakunin2004947,PhysRevLett.92.216806,
springerlink:10.1007/s10948-005-2144-x} and ab-initio LSDA
calculations \cite{wierzbowska_different_2004} indicate the
existence of an acceptor level or bound hybridized Mn pd-state at
$E_b \approx 112.4$ meV above the valence band.
This impurity state is absent for both values of
J$_{pd}$S = $\{1.2,3\}$ eV in the limit of $x\rightarrow 0$. In order to recover the correct Zeeman-splitting and impurity acceptor level energy
using J$_{pd}$S = 3 eV, one has to include a finite additional impurity potential scattering term V (see Eq.(\ref{eq:hamiltonian})) \cite{me}. Within a single-band model this term has been shown to be a crucial ingredient to understand magnetism and transport properties in III-V compounds \cite{0295-5075-92-4-47006,1367-2630-13-2-023002}.
Hence, our conclusion contradicts that of Ref.\cite{popescu_crossover_2007}, that a finite V is irrelevant.
Let us now discuss the nature of the magnetic couplings obtained within our multiband model.
The Mn-Mn exchange couplings $J_{ij}$ are calculated according to Eq.(\ref{eq:couplings}). In comparison to the MC studies of Ref.\cite{yildirim_large-scale_2007} our calculations are carried out for much larger systems (2048 lattice sites vs. 256) and the average over the disorder is performed over up to 800 vs. about 5 configurations in the MC simulations. Thus we expect our results to have much less finite size effects and reliable statistics.
For different values of J$_{pd}$S, $\bar{J}(R)\cdot R^3$ as a function of $R$ is shown in FIG. \ref{fig:3}., where
$\bar{J}(R) = \langle{J_{ij}}\rangle_{dis}$ and $R=\vert\textbf{r}_j-\textbf{r}_i\vert$.
Our results show clearly a major difference in the range, magnitude and nature of the
effective exchange couplings. For small values of J$_{pd}$S=$\{1.2, 3\}$ eV $\bar{J}(R)$ exhibits undamped long range RKKY oscillations. Hence, the couplings obtained for these values of J$_{pd}$ are inconsistent with those calculated from first principle studies for which RKKY oscillations are absent (see Ref.\cite{RevModPhys.82.1633}). This disagreement indicates that either the choice of parameters used to describe the physics of GaMnAs is incorrect or the model itself is inappropriate.
 In contrast, in the strong coupling regime J$_{pd}$S = $\{8,16\}$ eV, the exchange integrals appear to be of short-range ferromagnetic type. This regime can be seen as a precursor of the double-exchange regime, J$_{pd}$S = $\infty$. Both the RKKY and exponentially damped nature of the exchange  couplings can be clearly seen in FIG. \ref{fig:4}.
  Note that for J$_{pd}$S $\le$ 3 eV the Hamiltonian actually describes the physics of II-VI Mn-doped systems as, e.g. \mbox{CdMnTe} or \mbox{ZnMnSe}.
Thus, for J$_{pd}$S $\le$ 3 eV long-range ferromagnetic order is unlikely \cite{PhysRevB.73.024411}, because of high frustration. Instead one would expect a spin glass phase.
Finally, in FIG. \ref{fig:5}. we present the calculated Curie temperatures T$_C$ obtained by diagonalizing the Heisenberg Hamiltonian Eq.(\ref{eq:heisenberg}) within the SC-LRPA using different hole densities. For comparison both T$_C^{MC}$ (from Ref.\cite{yildirim_large-scale_2007}) and the mean-field virtual-crystal value (MF-VCA) T$_C^{MF}$ are shown. If we first focus on the largest hole density, p = 0.75$x$, we observe that T$_C^{MC}$ is surprisingly larger than T$_C^{MF}$. This is in contrast to the expectation that the MF value should overestimate the real critical temperature. Even in the region of small J$_{pd}$S ($\le$ 3eV), the MF value is zero whilst the MC value is always finite and relatively large. E.g. for J$_{pd}$S = 3 eV the MC value is about 500 K. Therefore one can question the reliability of the present MC calculations.
 On the other hand, the SC-LRPA T$_C$ is always much smaller than the MF values for the whole parameter range of J$_{pd}$S. Ferromagnetism is possible only for sufficiently large values of J$_{pd}$S. We find that the critical value is about 5 eV for p = 0.75$x$ and about 2 eV for p = 0.3$x$. Our calculated T$_C$ are approximately one order of magnitude smaller than those obtained from MC simulations. For example for J$_{pd}$S = 12 eV we have obtained 400 K within SC-LRPA whilst the MC simulations value is 2300 K. Note that for lower carrier density the critical temperatures are smaller but the ferromagnetism already appears for smaller values of J$_{pd}$S. This is expected since the effect of disorder is stronger at lower carrier concentration and RKKY oscillations will be suppressed for smaller values of J$_{pd}$S. Additionally, at lower density the period of the RKKY oscillations is larger thus the frustration effects weaken when carrier concentration is reduced.
Let us now explain the origin of the disagreement between our calculations and the MC simulations. The system considered in the MC study consists of only 256 lattice sites with typically only 20 localized spins and about 15 carriers/holes in the whole cluster. In addition the statistical average was carried out over about five configurations of disorder only.
As it was already pointed out in Ref.\cite{1367-2630-12-5-053042}, the T$_C$ is strongly size dependent and huge fluctuations of the critical temperatures distribution were observed for such small system sizes. For J$_{pd}$S $\le$ 4 eV, the smallness of the cluster used in the MC study does not resolve the asymptotic RKKY tail for J$_{pd}$S = 1.2 eV leading to finite and large Curie temperatures. In the large coupling regime, J$_{pd}$S $\ge$ 8 eV the T$_C^{MC}$ overestimates the real critical temperatures both due to insufficient statistical sampling and finite size effects. On the other hand, it is clear, that these essential numerical requirements are difficult to fulfill within standard MC calculations.
At last, one should underline, crucial differences between the two-step approach and MC simulations concerning the way T$_C$ is determined. Within SC-LRPA T$_C$ is directly calculated from a semi-analytical expression whilst in the present MC study it is extracted from the temperature variation of the magnetization curve. A more accurate way should be to use Binder cumulant in order to avoid additional errors, for the cost of significant computing time and memory needed for the finite size analysis. On the other hand, the SC-LRPA, allows the use of very large systems, typically of the order of $4\cdot(30)^3$ thus containing about 10$^4$ impurities.
\section{Conclusion}
In this work we have studied the magnetic properties of Ga$_{1-x}$Mn$_{x}$As including a realistic bandstructure of the host material. We have demonstrated that the parameters used in Ref.\cite{yildirim_large-scale_2007}, namely J$_{pd}$S = 1.2 eV and V = 0, cannot describe the magnetic properties in GaMnAs. The appearent agreement pointed out by those authors results from several levels of "approximations". Indeed, the direct comparison between our two-step approach and
  the recent MC simulations revealed several shortcomings. The MC study suffers from a) finite size effects, b) an insufficient statistical sampling and c) a less accurate and approximate procedure for the determination of the Curie temperature. It has been shown, that the critical temperatures T$_C^{MC}$ are largely overestimated for these reasons and even larger than mean-field values. We remark, that though the MC approach is essentially exact, the study of dilute magnetic systems has numerical requirements hard to fulfill by standard MC techniques. Furthermore we point out the necessity of including a finite V (impurity potential scattering term) in the multiband V-J model in order to describe the ferromagnetism in (Ga,Mn)As properly.

\end{document}